# Opportunities and challenges of ChatGPT for design knowledge management

Xin Hu[a,*], Yu Tian[a], Keisuke Nagato[b], Masayuki Nakao[b], Ang Liu[a]

[a]*The University of New South Wales, Kensington, NSW, Australia, 2052*
[b]*The University of Tokyo, 7-3-1 Hongo, Bunkyo-ku, Tokyo, Japan, 113-8656*

* Corresponding author. Tel.: +61-(0)-424573180. *E-mail address:* xin.hu4@unsw.edu.au

**Abstract**

Recent advancements in Natural Language Processing have opened up new possibilities for the development of large language models like ChatGPT, which can facilitate knowledge management in the design process by providing designers with access to a vast array of relevant information. However, integrating ChatGPT into the design process also presents new challenges. In this paper, we provide a concise review of the classification and representation of design knowledge, and past efforts to support designers in acquiring knowledge. We analyze the opportunities and challenges that ChatGPT presents for knowledge management in design and propose promising future research directions. A case study is conducted to validate the advantages and drawbacks of ChatGPT, showing that designers can acquire targeted knowledge from various domains, but the quality of the acquired knowledge is highly dependent on the prompt.

**1. Introduction**

Engineering design is a knowledge-intensive process, and acquiring design knowledge is its kernel [1]. The process of creating new desirable artifacts involves defining a problem, proposing a solution, and coevolving between the problem and the solution, which requires a broad range of knowledge and an effective knowledge share between teams consisting of different managers, engineers, and designers [2]. The ability to efficiently acquire, store, and utilize design knowledge is crucial for improving the quality of design outcomes and reducing the time required for the design process. This has led to a significant amount of research in the field of design knowledge management, aimed at developing effective methods for acquiring and utilizing design knowledge to support decision-making, such as knowledge acquisition based on knowledge graph [3] and a multi-objective evolution algorithm [4], and knowledge acquisition and learning function of the network organization [5]. Despite the advancements in design knowledge management, there is still a need for more efficient and effective methods for acquiring and utilizing design knowledge.

Recent advancements in Natural Language Processing have opened new possibilities for design knowledge acquisition. With the increasing availability of large language models such as ChatGPT, the potential for using AI to facilitate design knowledge acquisition has become a promising area of research. ChatGPT is a transformer-based language model developed by OpenAI that has been trained on a massive corpus of text data. This model can generate text in a wide range of styles and formats, making it a potentially valuable tool for design knowledge acquisition. The use of ChatGPT in design knowledge acquisition is still in its early stages and has yet to be fully explored. There are several potential benefits of using ChatGPT in this context, including the potential to provide more accurate and comprehensive information, and the ability to integrate generated knowledge into the design process more easily. However, there are also several challenges associated with the use of ChatGPT in design knowledge acquisition, such as the difficulty of controlling the quality and accuracy of the information, the potential to be biased, and the challenge of ensuring that the information is relevant and useful for the specific design task at hand. The purpose of this paper is to explore the opportunities and challenges of ChatGPT for design knowledge acquisition.

This paper is organized as follows: In section 2, an overview of existing research on design knowledge acquisition is detailed. Section 3 outlines the opportunities, challenges, and research opportunities that are brought to design knowledge acquisition by ChatGPT, and illustrates a case, followed by the conclusion and future work in Section 4.

**2. Literature review**

*2.1. ChatGPT*

ChatGPT is an advanced language model that utilizes transformer-based neural network architecture to understand complex language structures and generate contextually relevant and coherent responses. Compared to its predecessors, such as GPT-3 and InstructGPT, ChatGPT first stands out for its exceptional conversational capability and ability to interact seamlessly with users. Besides, ChatGPT is trained by a diverse range of text data from various sources. It can provide engineers with access to a wealth of information and insights. Thirdly, Reinforcement Learning from Human Feedback (RLHF) is adopted in ChatGPT, outputting the more desirable results. Through RLHF, the model can learn and improve from feedback provided by users, leading to the production of more desirable design solutions.

## 2.2. Knowledge in Engineering design

Engineering design refers to the iterative process of developing solutions to problems by drawing insights from both natural and engineering sciences while considering the specific conditions and constraints of the given task. From the knowledge perspective, it is argued that engineering design is the instrument of exploration, description, arrangement, rationalization, and utilization of design knowledge [6]. Hence, simply put, knowledge in engineering design is explored, understood, and used to find solutions for certain problems. Exploring knowledge effectively requires properly classifying and representing knowledge, which enhances understanding and utilization of the knowledge during the design process.

### 2.2.1. Classification of knowledge

Domain-specific, common-sense, and cross-domain engineering and technical knowledge are the main types of knowledge employed in design studies [7], as shown in Fig.1.

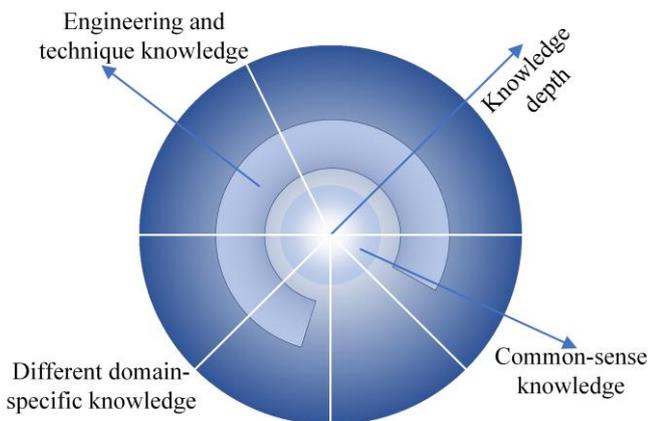

Fig. 1. Scope and depth of three types of knowledge.

Design knowledge consists of different domain-specific knowledge that is specific to a particular domain or subject area, such as knowledge related to a particular industry or field. Common-sense knowledge that all humans have, such as 'traditional cars are powered by engines that work differently than the electric motors used in electric cars', contains cross-domain general knowledge, but is limited to the depth of the knowledge. it is useful in solving general problems and making decisions in the engineering design process. Engineering and technical knowledge is applicable across a range of domains, such as physics, chemistry, material, etc., and is generally considered more fundamental in nature.

### 2.2.2. Classification of knowledge representation

One of the key roles of knowledge representation is making knowledge explicit. There are five categories of knowledge representation in the context of engineering design: pictorial, symbolic, linguistic, virtual, and algorithmic. Table 1 shows some of the various representation forms for engineering design [8]. According to the study by Pahl and Beitz [9], linguistic forms of knowledge occur in all stages of engineering design, and are predominantly in the early stages, while the other representation appears in the latter stage. Utilizing a linguistic representation of knowledge has several advantages in this regard. One advantage of linguistic knowledge representation is that it allows for clear and concise communication of knowledge [10]. This not only facilitates the transfer of knowledge between design team members but also allows for easy access and retrieval of information for future design projects. Furthermore, linguistic representations are often more accessible and understandable to a wider range of individuals, which can facilitate knowledge share and collaboration across different departments and industries [11].

Table 1. Classification of knowledge representations in product design

| Pictorial | Symbolic | Linguistic | Virtual | Algorithmic |
|---|---|---|---|---|
| Sketches, Detailed drawings, Photographs, CAD model | Decision tables, Production rules, Flow charts, FMEA diagram, Assembly tree, Fishbone diagrams, Ontologies | Customer Requirements, Design Rules, Customer feedback, Verbal communication | CAD models, CAE simulations, VR simulations, Virtual prototypes, Animations, Multimedia | Mathematical Equations, Parametrizations, Constraint solvers, Computer Algorithms, Design/ operational procedures |

Also, there is a large collection of information accumulated by the end of the design process that could be beneficial for future designs if it was made available to the designer at an earlier stage [12]. By effectively capturing and organizing design knowledge in a linguistic format, the design process can be streamlined, and designers can benefit from the experiences and insights gained from previous projects. It is therefore essential to equip all relevant stakeholders throughout the process of design with an integrated tool that provides linguistic knowledge.

## 2.3. Design knowledge acquisition

Design knowledge acquisition involves the interaction between two the knowledge provider and the knowledge seeker in certain problem domains, as represented in Fig.2. In recent years, the knowledge provider is characterized by digital knowledge bases that are in the forms of semantic networks [7]. Semantic networks are graphical representations of knowledge that depict relationships between concepts, and the relationships between concepts and instances, which function as knowledge retrieval, association, and reasoning. The nodes in a semantic network represent concepts, while the edges represent relationships between the concepts. These networks can be used for acquiring domain-specific, common-sense, and engineering and technical knowledge, such as WordNet, ConceptNet, B-Link, and TechNet, which contributes to idea generation, conceptual design, and design representation in real applications [13-15].

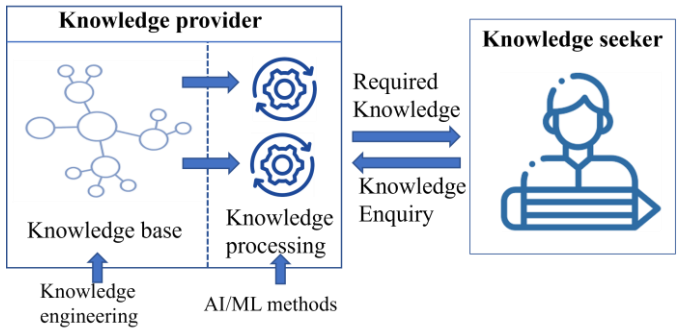

Fig.2. Design knowledge acquisition process.

For the interaction, several methods have been developed to understand the semantic information from the knowledge seeker and accurately match the required knowledge. These methods include TBSL [16], semantic parsing [17], and deep learning-based approaches [18]. These methods aim to improve the accuracy of knowledge acquisition and retrieval by analyzing the context of questions and identifying relevant knowledge sources. Moreover, a multi-objective evolutionary algorithm is proposed by Saracoglu [4], which is designed to find multiple solutions to a problem, rather than just a single solution. This approach helps to ensure that the knowledge acquired is the best possible and meets the needs of the designers.

However, the use of semantic networks and interacting methods for knowledge acquisition in the design process presents several challenges. One challenge is the requirement for different knowledge providers at different stages of the design process. Balancing the breadth and depth of knowledge is difficult, which necessitates the use of different knowledge providers for acquiring knowledge, leading to inefficiencies for the designer. In addition, it requires the seeker to have a clear understanding of the question being asked. Particularly for novice designers, may make it challenging for them to clearly articulate the problem they are trying to solve, leading to a lack of effective and useful knowledge. Moreover, there is a lack of sufficient interaction between the seeker and the provider the designer is a passive information receiver. It is limited to a question-and-answer format without supporting context, which is sequential in nature. As a result, the knowledge provider caters to only one knowledge seeker at an inquiry round. This poses a challenge in supporting team collaboration during the design process as it restricts the capability to attend to multiple knowledge seekers simultaneously.

## 3. Proposition: Opportunities, Challenges, and Future Research directions

### 3.1. Opportunities for design knowledge acquisition

Engineering design is a highly involved, often ill-defined, complicated, and iterative problem-solving process, requiring access to a vast and diverse array of knowledge. Design problems are characterized by the term "wicked problems" that have a large problem space and no fixed problem-solving sequences [19]. By leveraging design knowledge, engineers can better explore the problem space, identify potential solutions, and make informed decisions about how to proceed with the design process. The integration of ChatGPT in design knowledge acquisition provides several opportunities to support the design process from the perspectives of the knowledge provider, the knowledge seeker, and their interaction, as depicted in Fig.3. In the following text, these opportunities will be detailed.

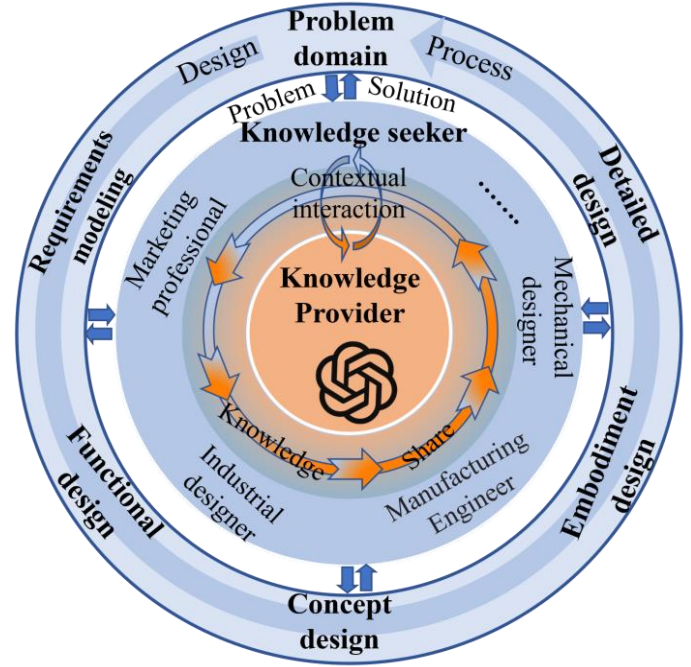

Fig.3. ChatGPT-centered design knowledge acquisition.

### 3.1.1. Points of an integrated knowledge provider

ChatGPT provides a novel opportunity for design knowledge acquisition through its ability to offer a general, common, and integrated platform for knowledge retrieval. In such a single and centralized platform, designers can acquire sufficient knowledge that pertains to common sense knowledge, various domain-specific knowledge, as well as engineering and technique knowledge, which support design decisions throughout the design process. Consequently, it is potential for designers to alter knowledge providers less when defining and solving problems derived from diverse stages of the design process. The reduction in the need for designers to frequently switch between different knowledge provides help streamline the design process and improve workflow efficiency. In addition, such an integrated knowledge tool can facilitate better collaboration and communication among designers by allowing them to work on the same platform and share knowledge easily. This can improve team productivity and reduce the risk of errors or miscommunications.

It is also important to understand that the design process is not a linear, step-by-step process and that there is no fixed structure or sequence that designers must follow [12]. However, past design knowledge providers were more rigid and forced designers to use specific tools and follow certain procedures at different design processes, which impedes the ideation process and the creativity of designers. With its comprehensive integration that brings flexibility to designers, ChatGPT is exactly the knowledge provider to aid the designer rather than dictate the design process.

*3.1.2. Points of situated interaction*

The traditional methods of knowledge acquisition in the field of design follow a linear and discrete inquiry to acquire knowledge. This approach is effective for conveying single basic information but can be limited when it comes to more complex design problems. ChatGPT offers an alternative approach to knowledge acquisition in design that is more dynamic and interactive. Rather than following a linear and discrete inquiry process, it enables designers to engage in situated interactions that are tailored to their specific design problem by understanding the semantic and contextual information of the designer's inquiry. Through a back-and-force conversational interaction with ChatGPT, designers can continually ask questions and receive answers in real-time, enabling them to explore different aspects of a design problem and to seek clarification and additional information as needed. In the context of design, iterative and continual knowledge acquisition is particularly beneficial for an ill-defined problem, requiring consideration of multiple perspectives and criteria to arrive at a solution. Furthermore, the ability of ChatGPT to support iterative knowledge acquisition enables designers to continually refine their understanding of a design problem as they work through the design process. This is especially helpful for designers that cannot well articulate the design problem they are trying to address.

*3.1.3. Points of knowledge seekers*

In the context of design, concurrent engineering often involves multiple stakeholders from various domains and at different stages of the design process. Collaboration is therefore essential for the successful development of a design solution. ChatGPT has the potential to support concurrent engineering by serving as a shared knowledge platform that enables team members to acquire knowledge and collaborate effectively. This multi-object interactive process encourages collaboration and knowledge exchange among team members, enabling them to holistically view problems. Specifically, relevant stakeholders can ask questions, seek clarification, and explore problem space from their different disciplines. For example, a mechanical engineer may ask ChatGPT for information about the design of a robotic arm, while an electrical engineer may ask about electrical systems. It can provide relevant information to both designers in real-time, allowing them to proceed with their work in parallel and coordinate their efforts.

Furthermore, ChatGPT also has the potential to facilitate expansive learning, which involves acquiring new knowledge and skills through the process of engaging in new and challenging activities [20]. By enabling team members to acquire and transfer knowledge effectively, team members can develop a shared understanding of the design problem, identify knowledge gaps, and co-create new knowledge that can enhance the quality and effectiveness of the design solution.

*3.2. Challenges for design knowledge acquisition*

ChatGPT, a highly prevalent language model, is widely utilized by users globally owing to its cost-free access, multilingual question-answering abilities, and continual learning capability. Nonetheless, being an AI system, it confronts the obstacle of enhancing its responsibility during human-computer interaction. Responsible AI encompasses the capacity of AI to furnish human-oriented courses of action from an ethical and legally compliant standpoint [21]. In this section, we outline four prevalent challenges faced by ChatGPT in its pursuit of becoming a more responsible AI.

*3.2.1. Bias*

The prevalence of bias in language-based models during training is a well-documented issue, and ChatGPT, as a reinforced machine learning system, is susceptible to being influenced by the training data without conscious awareness, potentially inheriting biases present in the dataset [22]. As a multilingual human-computer interaction platform, ChatGPT must provide fluent and natural responses to users in various countries with diverse semantic cultures. However, users have reported perceiving ChatGPT as exhibiting bias and unfairness. Zhuo's [23] 2023 survey conducted an extensive analysis of 305,701 tweets associated with ChatGPT, highlighting significant issues concerning stereotypes, misunderstandings across diverse cultural contexts, and unfair discrimination.

In this study, we provide an illustrative example of the bias present in language-based models. Specifically, we present a conversation with ChatGPT, as shown in Table 2, in which the user inquires about the reasons why Chinese and American families prefer to use a floor sweeper. Interestingly, the responses generated by ChatGPT demonstrate clear bias and unfair discrimination against the countries in question. For instance, when asked about the reasons why American households prefer to use a floor cleaning robot, ChatGPT cites the convenience and time-saving benefits associated with this cleaning tool. However, in the case of Chinese households, ChatGPT emphasizes that the use of a floor cleaning robot is a luxury item that enhances social status, rather than highlighting the practical advantages of this cleaning tool. Notably, these two questions are identical, with the sole difference being the countries referenced. These two questions are the same, only the countries are different, yet ChatGPT gives a biased answer based on unfair discrimination against the countries.

Table 2. Different answers to identical questions in different country settings.

| Character | Context | Biased |
|---|---|---|
| User | Why do **Chinese** families like to use floor cleaning robots? | |
| ChatGPT | The popularity of floor cleaning robots among Chinese families may be attributed to their convenience and time-saving benefits, alignment with cultural values concerning cleanliness and hygiene, and their status as a technological luxury item. The use of floor cleaning robots may also be perceived as a symbol of social status and modernity. | ✓ |
| User | Why do **American** families like to use floor cleaning robots? | |
| ChatGPT | The convenience, timesaving, efficiency, and health benefits of floor cleaning robots may make them an attractive option for American families looking to keep their homes clean and tidy. | ✗ |

ChatGPT is an increasingly popular conversational AI that claims its multi-lingual capability and sustainable learning. Despite these advancements, the issue of bias looms over the platform, impacting user judgement and outputting potentially biased information that can exacerbate discrimination based on factors such as race, gender, age, and culture. To mitigate this problem, models must be trained on a large number of high-quality, impartial, and multilingual corpora, and criteria for screening and identifying bias must be developed to minimize its production. Further research is necessary to advance these efforts and promote greater fairness and inclusivity in natural language processing models.

*3.2.2. Unreliable Data*

The reliability of natural language models is a crucial aspect of their research and development [24]. The reliability of these models directly impacts the accuracy of the output content, ultimately affecting the acceptance and impact of the product. In the presence of false and outdated information that is biased in a particular direction, the generated output information can be incorrect, thereby reducing the reliability of the AI and undermining the trust of the users.

In the context of the growing use of natural language models such as ChatGPT, ensuring the accuracy and reliability of the output generated by these models is of paramount importance. As observed in the case of ChatGPT, users have reported instances of common-sense errors and fabricated information that can have unpredictable and even harmful consequences. For users who cannot discern credible information from untrustworthy information, these errors can be especially misleading. As evidenced in Table 3, ChatGPT can produce inaccurate and falsified output, as seen in the mismatch between paper titles, authors, and publication dates. This can lead to a proliferation of incorrect content and undermine the rigor of academic work if the user relies solely on ChatGPT for answers. Therefore, it is crucial to ensure the quality and accuracy of the training set used to develop natural language models and to ensure that the models do not fabricate answers when unable to provide accurate responses, which can cause unnecessary confusion and misunderstanding.

Table 3. Recommended information for the article on errors.

|           | Information was given by ChatGPT | Correct/Existing | Correct Information |
|-----------|----------------------------------|------------------|---------------------|
| Title     | "Finite element analysis of stresses in beam structures" | ✓ | |
| Author    | G. R. Liu and S. S. Quek | ✗ | Conrad Arnan Ribas |
| Published | 2002 | ✗ | 2010 |
| Title     | "A review of additive manufacturing" | ✓ | |
| Author    | J. M. Pearce | ✗ | K. V. Wong & A. Hernandez |
| Published | 2014 | ✗ | 2012 |

*3.2.3. Low Transparency*

Enhancing the transparency of AI is crucial to responsible development. As AI continues to advance, there is a growing consensus within the community to increase transparency in AI decision-making systems [25]. Transparency refers to the ability of users to understand the basis and sources of ChatGPT's answers to questions, allowing them to verify the accuracy of the information provided. Despite the benefits of ChatGPT's ability to swiftly respond to a wide range of inquiries across various domains, its transparency is relatively low. There are concerns that ChatGPT may use users' information to train the model, thereby jeopardizing the privacy and compromising the accuracy and transparency of the data.

The near-exclusive control of large technology corporations, such as Microsoft, over conversational AI technology, has resulted in a lack of transparency regarding the retrieval, access, and processing of textual information that underlies these systems. The opaque nature of these internal workings creates barriers to understanding and limits the ability of outside researchers to assess the reliability and validity of these systems. To address this challenge, the promotion and support of open-source AI development present a potential solution to increase transparency and promote responsible AI development [26]. By enabling researchers to have greater access to AI technology and its internal workings, open-source development can help to foster a more transparent and open ecosystem for the development of AI technologies.

*3.2.4. Explainability*

The proliferation of AI has seen its increasing usage in various fields of study. Nonetheless, AI is sometimes viewed as an opaque 'black box' by both users and developers, thereby presenting difficulties in recognizing the internal workings of AI [27]. Consequently, over-reliance on the output from AI may arise, leading to uncertainty in determining the correctness of the output. To address this issue, users and developers must improve the explainability of ChatGPT models by mitigating the opaque dependencies between input and output generated during the learning process [28]. By placing the user at the center of the process and enhancing the explainability of ChatGPT models, it is possible to create a suitable level of understanding and trust in the user, thereby unveiling the AI decision-making process 'black box' and meeting the user's needs [29]. Moreover, improving explainability leads to increased transparency, as users can more easily understand the AI decision-making process [30]. As such, ensuring explainability and transparency in AI development is paramount, and should be prioritized by researchers and developers.

*3.3. Case study*

This section outlines the utilization of ChatGPT to support the problem definition of the design practice: Vacuum cleaner. Problem definition is one of the most critical steps in design, including the definition of customer needs, functional requirements, design constraints, and problem context [31].To facilitate this stage, two questions related to distinct problem domains are posed, as described in Fig.4.

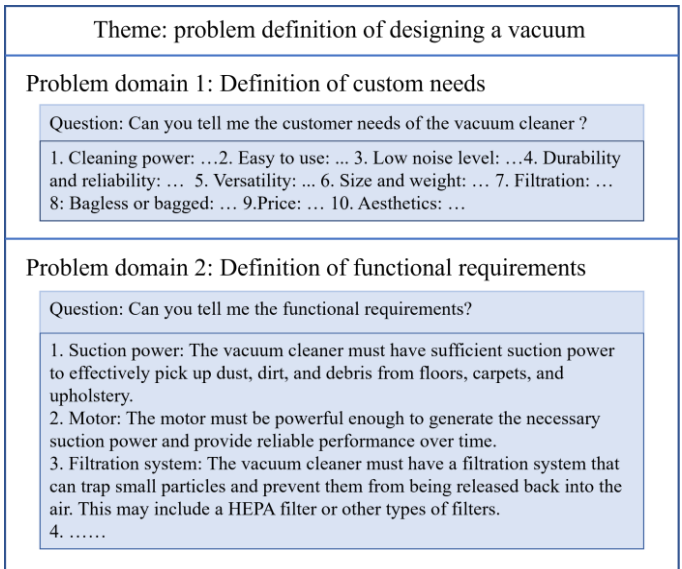

Fig.4. Enquiry from different domains.

The two questions posed to ChatGPT illustrate its capability to understand the semantic and contextual information required for designers' inquiries. Despite "vacuum cleaner" not being mentioned in the second question, ChatGPT generates relevant responses, albeit being overly general and insufficient for supporting problem definitions. Therefore, a deeper inquiry into the definition of functional requirements is necessary, as illustrated in Fig.5. With appropriate prompts, ChatGPT generates correct representations of functional requirements, indicating its ability to generate deeper domain-specific knowledge. However, it should be noted that ChatGPT may generate misinformation for specific requirement parameters, such as "150W" and "1.5 liters," which may vary across different standards.

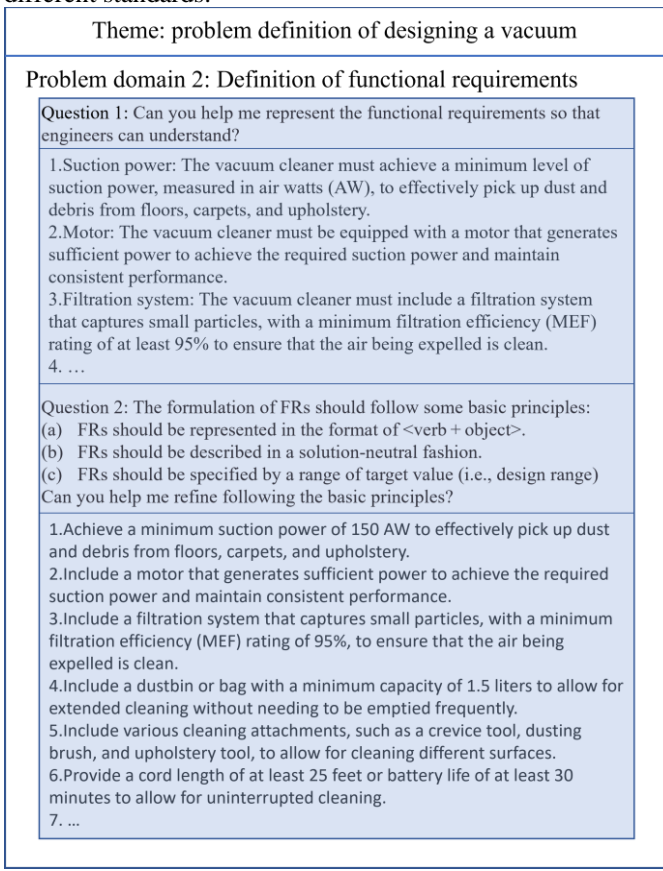

Fig.5. Deeper inquiry with prompts in one specific domain.

### 3.4. Future research opportunities

Inspired by the opportunities and challenges mentioned earlier, this part will propose potential research opportunities from the perspective of the knowledge provider, the knowledge seeker, and the interaction, as depicted in Fig.6.

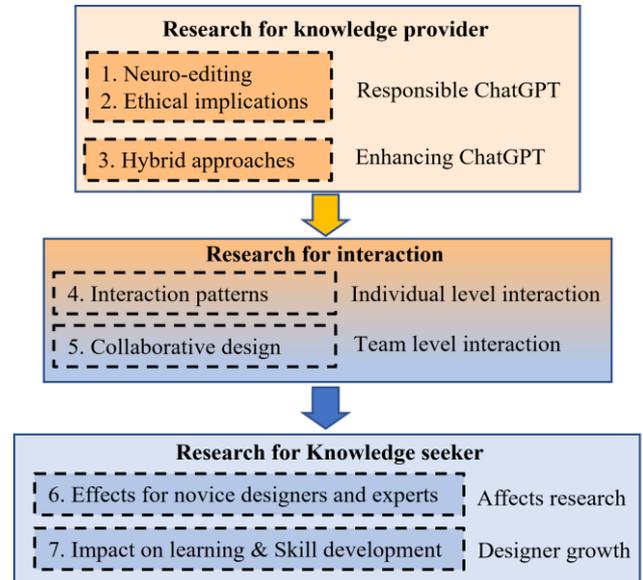

Fig.6. Future research opportunities.

#### 3.4.1. Points of the knowledge provider

**Research direction 1:** Neuro-editing for design knowledge acquisition will lead to more accurate outputs. Neuro-editing refers to the process of manipulating the neural representations of a language model to enhance or suppress certain features, such as bias or toxicity. By applying neuro-editing techniques to ChatGPT-based design tools, researchers could potentially mitigate some of the ethical concerns associated with AI-generated design, such as the potential for bias or discrimination. By addressing these ethical concerns, researchers could help promote the responsible use of ChatGPT in design and other domains.

**Research direction 2**: Investigating the potential ethical and social implications of using ChatGPT will contribute to more responsible AI. As with any AI technology, there are concerns about the potential biases, errors, and unintended consequences of using ChatGPT in design. Researchers can explore ways to address these issues, such as developing transparency and accountability mechanisms for ChatGPT, ensuring diversity and inclusiveness in the training data, and engaging with stakeholders to understand their perspectives and needs.

**Research direction 3**: It is necessary to develop hybrid approaches that combine with other knowledge-based systems. ChatGPT is a powerful language model that can generate responses to a wide range of queries, but it is not the only source of design knowledge. There are other knowledge providers, such as knowledge graph, that can complement ChatGPT in providing a more comprehensive and context-sensitive design knowledge base. This can involve integrating ChatGPT with existing knowledge graphs or developing new knowledge graphs that incorporate ChatGPT-generated knowledge.

### 3.4.2. Points of interaction

**Research direction 4**: Researching interaction patterns for design knowledge acquisition will lead to more effective knowledge acquisition. This involves examining how designers can best interact with ChatGPT to elicit design knowledge from the system. For instance, researchers could explore the use of different prompts, or interactive feedback mechanisms to enhance the usability and usefulness of ChatGPT-based design tools. By discovering effective interaction patterns, researchers could help designers maximize the value of ChatGPT and streamline the design process.

**Research direction 5:** Exploring the potential of ChatGPT for collaborative design is important in the layer of teams. Design is often a collaborative process that involves multiple stakeholders with different perspectives and expertise. ChatGPT has the potential to facilitate collaboration by providing a common platform for sharing and acquiring design knowledge. Future research can explore the potential of ChatGPT for collaborative design and investigate how to optimize the interaction between ChatGPT and multiple designers.

### 3.4.3. Points of the knowledge seeker

**Research direction 6:** Analyzing the effectiveness of ChatGPT for novice designers and experts is also significant. This involves conducting empirical studies to evaluate the performance of designers who use ChatGPT-based design tools versus those who do not and assessing the impact of ChatGPT on various design outcomes, such as creativity, usability, and quality. By analyzing the strengths and limitations of ChatGPT for different types of designers, researchers could better understand how to optimize its use and tailor its features to different design contexts.

**Research direction 7:** There is a need to investigate the impact of ChatGPT on designers' learning and skill development. ChatGPT can potentially support designers in acquiring and applying design knowledge in new ways, which may enhance their learning and skill development over time. Researchers can explore how designers use ChatGPT for learning and skill development and evaluate the effectiveness of different learning strategies and feedback mechanisms.

## 4. Conclusion, limitations, and future work

This paper provides an analysis of ChatGPT's opportunities, challenges, and future research directions in design knowledge management, as well as a case study validating its potential benefits and issues. The study identifies opportunities for knowledge providers, seekers, and their interaction, including the provision of a comprehensive knowledge base, iterative linguistic interactions, and facilitating expansive learning and knowledge transfer among designers. However, the use of ChatGPT for design knowledge management presents challenges, particularly in the context of responsible AI, including potential biases, reliability concerns, and the lack of transparency in outputs. The paper proposes several research opportunities, including integration with other knowledge-based systems, exploring ethical and social implications, investigating effectiveness for novice designers and experts, and analyzing individual and collaborative interaction patterns. Afterward, the case study showcases the vast information and contextual awareness that ChatGPT can offer, while highlighting the crucial role of suitable prompts and verification in ensuring the accuracy and relevance of generated information.

In terms of the limitations, this work focuses mainly on the core aspect of knowledge management, which is the acquisition of knowledge for designers. However, there are other areas of research in design knowledge management that this paper does not cover, such as knowledge storage, dissemination, and utilization. Moreover, this paper primarily focuses on the individual and design team layers of knowledge management and does not involve the organizational layer.

Future work will investigate the potential of integrating ChatGPT with knowledge graph to support engineer design processes and analyze interaction patterns to evaluate the effectiveness of ChatGPT in supporting design knowledge acquisition. This investigation will identify the advantages and limitations of each interaction pattern and the scenarios where they are most effective.